\documentstyle[epsf]{article}
\newcommand{\etal}{{\it et al.}}

\newenvironment{refs}{\section*{References} \begin{list}{}{
         \setlength{\parsep}{0pt} \setlength{\itemsep}{0pt}
         \setlength{\leftmargin}{0pt} } }{ \end{list} }
\title{A Search for Ultra-High Energy Counterparts to Gamma-Ray Bursts}
\author{S.P. Plunkett, M. Delaney, B. McBreen, K.J. Hurley and C.T. O'Sullivan}
\date{astro-ph/9508083: presented at 29 ESLAB Symposium, April 1995}
\begin{document}

\maketitle
\begin{abstract}
  A small air shower array operating over many years has been used to
  search for ultra-high energy (UHE) gamma radiation ($\geq 50$ TeV)
  associated with gamma-ray bursts (GRBs) detected by the BATSE
  instrument on the Compton Gamma-Ray Observatory (CGRO).  Upper
  limits for a one minute interval after each burst are presented for
  seven GRBs located with zenith angles $\theta < 20^{\circ}$. A
  $4.3\sigma$ excess over background was observed between 10 and 20
  minutes following the onset of a GRB on 11 May 1991. The confidence
  level that this is due to a real effect and not a background
  fluctuation is 99.8\%. If this effect is real then cosmological
  models are excluded for this burst because of absorption of UHE
  gamma rays by the intergalactic radiation fields.
\end{abstract}

\section{Introduction}

Despite the large number of GRBs detected, the sources responsible for
this extraordinary phenomenon remain unidentified (Harding, 1994;
Hurley, 1994).  To date, the majority of GRB counterpart searches have
been carried out at low energies (e.g.\ Schaefer, 1994), with few
attempts to search for a UHE component (e.g.\ Alexandreas \etal, 1994;
Borione \etal, 1993; Connaughton \etal, 1993; Vallania \etal, 1993).
Kazanas and Ellison (1986) have predicted that proton acceleration due
to diffuse shock acceleration in the atmosphere of neutron stars could
produce UHE emission, while the lightning model of McBreen \etal\
(1994) and the fireball model of M\'{e}sz\'{a}ros and Rees (1993)
suggest a similar result.  Delayed high energy gamma ray emission,
including an 18 GeV photon detected 1.5 hours after the onset of
GRB940217, has been detected by the EGRET instrument onboard CGRO
(Dingus \etal, 1994; Hurley \etal, 1994). Models predicting such
emission were subsequently proposed by M\'{e}sz\'{a}ros and Rees (1994)
and Katz (1994).

A positive detection of UHE emission would indicate an upper limit to
the distance to GRB sources, since photons of such energy undergo pair
production with soft photons of the intergalactic radiation fields.
Absorption by the cosmic microwave background (CMB) would impose a
distance constraint of $\sim\! 100$ Mpc for 100 TeV photons (Gould and
Schreder 1966; Jelley 1966), though greater attenuation by the
intergalactic infrared radiation field (IIRF) may reduce this to
$\sim\! 10$ Mpc (Stecker \etal, 1994).

A small air shower array (Plunkett \etal, 1991a; 1991b) has been in
operation near sea level at University College, Dublin, since early
1989. Briefly, the array consists of eight plastic scintillation
counters of area 1 m$^{2}$, which sample air showers over a collection
area approximately $1.2 \times 10^{3}$ m$^{2}$ above a threshold
primary particle energy of about 50 TeV for showers close to the
zenith and a median primary particle energy of about 100 TeV. The
angular resolution of the array is 3.1$^{\circ}$ when all eight
detectors are triggered, and about 4.5$^{\circ}$ when three are
triggered. The latter resolution was used because showers triggering
at least three detectors were accepted for analysis.

The long term stability of the array was checked by the analysis of
$10^{4}$ hours of accumulated data. Over this period, the number of
air showers recorded over intervals of one minute were compared to a
measure of the mean rate of arrival of showers, determined over a
period of $\sim\! 6$ hours to account for variations imposed by
changes in atmospheric pressure. The mean shower arrival rate was
typically 35 min$^{-1}$, allowing the normal approximation to the
Poisson distribution to be applied to this analysis.  The statistical
significance of any positive excesses was computed by the maximum
likelihood method of Li and Ma (1983).  Both the observed and expected
number of one minute time intervals which registered an excess of air
showers with significances in the ranges $3.5\!-\!4\sigma$,
$4\!-\!4.5\sigma$, $4.5\!-\!5\sigma$ and $>\!5\sigma$, were shown to
give good agreement, thus confirming the stability of the array.

\section{Data Analysis}
The BATSE GRB catalogues (Fishman \etal, 1994) were searched for
bursts with a reported location within the field of view of the array
and a Poisson positional uncertainty of $<\! 5^{\circ}$, not including
systematic errors.  This search yielded a sample of seven GRBs with a
zenith angle $\theta\! <\! 20^{\circ}$.

The search for UHE emission was conducted for a period of up to three
hours before and after each burst, while the reported burst location
was at $\theta\! <\! 20^{\circ}$.  This period was divided into
non-overlapping time intervals of ten minutes duration, to search for
weak sustained emission, and an additional single interval of one
minute duration after the GRB onset time.  The number of showers from
the burst direction was computed using a optimally sized circular
source bin, which is 1.6 times the angular resolution of the array
(Alexandreas \etal, 1993), of radius $7^{\circ}$ centred on the
reported location of the burst.  The expected background was computed
using up to eight non-overlapping bins, identical in size to the
source bin, separated by $14^{\circ}/\cos\delta$ in right ascension
$\alpha$, but having the same declination $\delta$ as the GRB
location.  These bins followed the same path on the sky as the source
bin, but at different times, and ensured a proper account of the
variation in shower arrival rate with zenith angle for all bins.

Upper limits to the photon flux were computed by the method of Gatto
\etal\ (1988) over a period of one minute after the GRB onset.  In the
case of only one burst, GRB910511, a significant delayed excess of
$>\! 3 \sigma$ was found at the reported location, prompting a more
thorough search for the location of an optimum excess.  The search
process was repeated 28 times with the source and background bins
shifted first in increments of $3^{\circ}$ in $\alpha$ and $\delta$ to
determine within which quadrant the excess is optimised, and
subsequently in smaller steps down to $1^{\circ}$ to further localise
the optimum source bin.

\section{Results}
The analysis of the seven GRBs in the sample showed no evidence for
coincident UHE emission. Upper limits at the 95\% confidence level,
calculated over an interval of one minute after the GRB onset, are
listed in Table~\ref{limits}.

\begin{table}[t]
  \caption{\label{limits}Upper limits to the number of
    air showers detected by the
    array and the UHE photon flux at the 95\% confidence level for the GRBs
    listed.}
  \begin{center}
    \begin{tabular}{lllll}\hline
      GRB Date & Time & Altitude & Air Showers & Photon Flux \\
      (yymmdd) & (UT) & ($90^{\circ} - \theta$) & & (cm$^{-2}$ s$^{-1}$)\\
      \hline
      910511 & 02:11:47.720 & 81.9 & $<\!6$ & $<\!1.7 \times 10^{-8}$ \\
      910512 & 15:15:09.449 & 71.6 & $<\!8$ & $<\!2.2 \times 10^{-8}$ \\
      910718 & 02:24:57.230 & 75.9 & $<\!3$ & $<\!8.3 \times 10^{-9}$ \\
      911004 & 01:27:09.760 & 70.2 & $<\!5$ & $<\!1.4 \times 10^{-8}$ \\
      920320 & 12:18:58.879 & 73.6 & $<\!5$ & $<\!1.4 \times 10^{-8}$ \\
      921101 & 18:03:55.738 & 72.9 & $<\!5$ & $<\!1.4 \times 10^{-8}$ \\
      931220 & 15:50:01.000 & 75.3 & $<\!7$ & $<\!1.9 \times 10^{-8}$ \\
      \hline
    \end{tabular}
  \end{center}
\end{table}

The analysis of GRB910511 revealed an excess of events from the
reported location, between 10 and 20 minutes after the burst onset.
This excess was noted long before the report on delayed emission by
Hurley \etal\ (1994).  As determined by BATSE this was a burst of
duration 7.2 s with a peak photon flux of 4.7 cm$^{-2}$ s$^{-1}$,
located at $\alpha\! =\! 266.5^{\circ}$, $\delta\! =\! 57.2^{\circ}$,
and with a Poisson positional uncertainty of $3.1^{\circ}$.  This
location remained at $\theta\! <\! 20^{\circ}$ for 1.6 hours before
the burst and 3.1 hours afterwards.  A detailed search, described in
section two, revealed an optimum excess in a bin centred on the
location $\alpha\! =\! 269.5^{\circ}$, $\delta\! =\! 63.2^{\circ}$
with a positional error of about $3^{\circ}$.  The number of counts in
ten minute intervals from this source bin is shown in Fig.\ 1.  During
a ten minute interval beginning at 02:21:47 UT, a total of 36 air
showers were detected compared to an expected background of 14.6
showers. This corresponds to an excess of $21.4 \pm 6.2$ events (the
errors quoted are $\pm 1 \sigma$), giving a statistical significance
of $4.3 \sigma$ (Li and Ma 1983).  Upon examination, the arrival times
of these showers appear consistent with a random distribution.

\begin{figure}[t]
  \epsfxsize=\textwidth
  \epsffile{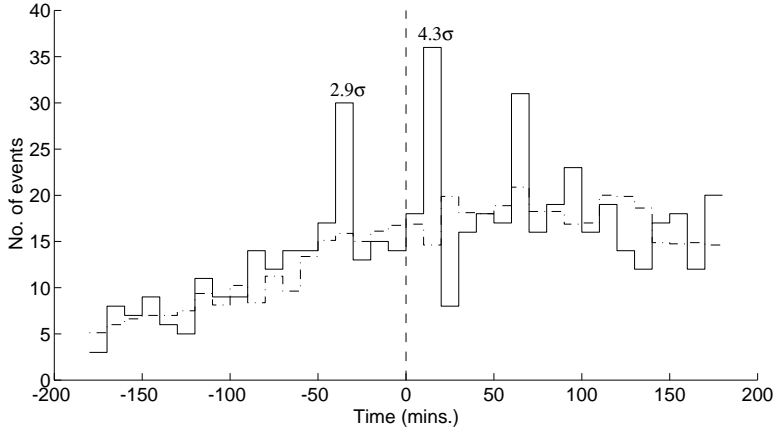}
  \caption{The solid line gives the number of air showers
    detected as a function of time from the optimum location
    within the BATSE uncertainty
    in the position of GRB910511. The expected background rate in each time
    interval is shown as a dashed line. Time is measured in minutes
    relative to the GRB onset at 02:11:47 UT. The $4.3 \sigma$ and an
    earlier $2.9 \sigma$ excess are shown.}
\end{figure}

The initial number of trials made in the analysis of all seven GRBs
was 150.  A further 62 trials were conducted during the overlapping
bin search for the optimum location of the delayed excess of air
showers in the interval 10 to 20 minutes after the onset of GRB910511,
and in examining the behaviour of the source over a six hour period at
this location (see Fig. 1).  This yields a total of 212 trials and
results in a confidence level of 99.8\% that this excess was caused by
an UHE burst from the direction of the GRB, which occurred more than
ten minutes earlier, and not by a background fluctuation.  The flux of
photons which is required to give the observed excess is $(5.9 \pm
1.7) \times 10^{-9}$ cm$^{-2}$ s$^{-1}$, assuming an efficiency of 0.5
for detection of air showers. Taking a typical primary particle energy
of 100 TeV for the showers detected by the array, this corresponds to
a fluence of $(9.5 \pm 2.8) \times 10^{-7}$ erg cm$^{-2}$ s$^{-1}$
averaged over the ten minute duration of the burst. An earlier excess
of showers appears in Fig. 1, though with a low significance of $2.9
\sigma$ this was rejected as a possible UHE detection.

\section{Conclusions}
For all seven of the GRBs in the sample, upper limits were obtained
for UHE emission in a one minute time interval after the GRB onset.
Evidence for delayed UHE emission accompanying GRB910511, in the
interval 10--20 minutes after the initial detection, has been
presented. If this effect is real an upper limit of 100 Mpc due to
absorption by the CMB is imposed on the distance to the source, which
might be further reduced to the order of 10 Mpc by the IIRF. It may
therefore be concluded that a non-cosmological origin is expected for
the source of GRB910511.  Given that a large number of number of
ingenious mechanisms have been proposed to account for the generation
of GRBs (Nemiroff, 1994), it may be considered possible that there are
contributions to the observed number of bursts from sources at both
cosmological and non-cosmological distances.

\section*{Acknowledgements}
{The authors are indebted to Catherine Handley for her able
assistance with the data analysis and to FORBAIRT for their support.}

\begin{refs}
\item Alexandreas, D.E. \etal, 1993, Nucl. Instrum. Meth. Phys. Res.,
  A328, 570.
\item Alexandreas, D.E. \etal, 1994, Astrophys. J., 426, L1.
\item Borione, A. \etal: 1993,
{\it Proc. 23rd ICRC}, Calgary, {\bf 1}, 57.
\item Connaughton, V. \etal: 1993,
{\it Proc. 23rd ICRC}, Calgary, {\bf 1}, 69.
\item Dingus, B.L. \etal: 1994,
{\it AIP Conf. Proc. 307}, AIP:New York, 22.
\item Fishman, G.J. \etal, 1994, Astrophys. J. Suppl., 92, 229.
\item Gatto, R. \etal, 1988, Phys. Lett., B204, 81.
\item Gould, R.J. and Schreder, G. \etal, 1966, Phys. Rev.
Lett., 16, 252.
\item Harding, A.: 1994, in C. Fichtel \etal\ (eds.)
{\it AIP Conf. Proc. 304}, AIP:New York, 30.
\item Hurley, K.C., 1994, Astrophys. J. Suppl., 90, 857.
\item Hurley, K.C. \etal, 1994, Nature, 372, 652.
\item Jelley, J.V. \etal, 1966, Phys. Rev. Lett., 16, 179.
\item Katz, J., 1994, Astrophys. J., 432, L27.
\item Kazanas, D. and Ellison, D.C., 1986, Adv. Space Res., 6, 81.
\item Li, T.P. and Ma, Y.Q., 1983, Astrophys. J., 272, 317.
\item M\'{e}sz\'{a}ros, P. and Rees, M.J., 1993, Astrophys. J., 418, L59.
\item M\'{e}sz\'{a}ros, P. and Rees, M.J., 1994, Mon. Not. Roy. Astr. Soc.,
269, L41.
\item McBreen, B. \etal, 1994, Mon. Not. Roy. Astr. Soc., 271, 662.
\item Nemiroff, R.J.: 1994,
{\it AIP Conf. Proc. 307}, AIP:New York, 730.
\item Plunkett, S. \etal, 1991a, Nucl. Instrum. Meth. Phys. Res., A300, 197.
\item Plunkett, S. \etal, 1991b, Proc. 22nd ICRC (Dublin), 1, 89.
\item Schaefer, B.E.: 1994,
{\it AIP Conf. Proc. 307}, AIP:New York, 382.
\item Stecker, F.W. \etal, 1994, Nature, 369, 294.
\item Vallania, P. \etal: 1993,
{\it Proc. 23rd ICRC}, Calgary, {\bf 1}, 61.
\end{refs}
\end{document}